\documentclass[prb,twocolumn,superscriptaddress,showpacs,amsmath,amssymb]{revtex4}
\usepackage{amsfonts}
\usepackage{bbm}
\usepackage{graphicx}% Include figure files
\usepackage{dcolumn}% Align table columns on decimal point
\usepackage{bm}% bold math

\begin{document}
\title{Specular Andreev reflection and its detection}

\author{Qiang Cheng}
\affiliation{School of Science, Qingdao University of Technology, Qingdao, Shandong 266520, China}
\affiliation{International Center for Quantum Materials, School of Physics, Peking University, Beijing 100871, China}

\author{Qing-Feng Sun}
\email[]{sunqf@pku.edu.cn}
\affiliation{International Center for Quantum Materials, School of Physics, Peking University, Beijing 100871, China}
\affiliation{Collaborative Innovation Center of Quantum Matter, Beijing 100871, China}
\affiliation{CAS Center for Excellence in Topological Quantum Computation, University of Chinese Academy of Sciences, Beijing 100190, China}

\begin{abstract}
We propose a universal method to detect the specular Andreev reflection taking the simple two-dimensional Weyl nodal-line semimetal$-$superconductor double-junction structure as an example. The quasiclassical quantization conditions are established for the energy levels of bound states formed in the middle semimetal along a closed path. The establishment of the conditions is completely based on the intrinsic character of the specularly reflected hole which has the same sign relation of its wave vector and group velocity with the incident electron. This brings about the periodic oscillation of conductance with the length of the middle semimetal, which is lack for the retro-Andreev reflected hole having the opposite sign relation with the incident electron. The positions of the conductance peaks and the oscillation period can be precisely predicted by the quantization conditions. Our detection method is irrespective of the details of the materials, which may promote the experimental detection of and further researches on the specular Andreev reflection as well as its applications in superconducting electronics.
\end{abstract}
\maketitle

\section{\label{sec1}Introduction}

Andreev reflection is the fundamental scattering process in the metal-superconductor (SC) heterojunctions, in which an incident electron from the metal is reflected as a hole at the metal-SC interface and a Cooper pair forms in the SC.\cite{Andreev} When bias is less than the superconducting gap, the Andreev reflection dominates the conductance of the metal-SC heterojunction\cite{Blonder,Sun1,ZPNiu}.
Furthermore, the Andreev reflection is also responsible for various of basic physical phenomena such as the Josephson effects\cite{Golubov}, the proximity effects\cite{Buzdin} and the odd frequency pairings\cite{Bergeret,Linder}. For a conventional metal (CM) attached by SC,
the reflected hole moves back along the trajectory of the incident electron. This type of Andreev reflection is called the retro-Andreev reflection (RAR). Recently, with the rise of quantum materials, the so-called specular Andreev reflection (SAR) is first predicted in graphene\cite{Beenakker}. In SAR process, an incident electron will be specularly reflected as the hole.
Up to now, the SAR is discovered and studied extensively in the monolayer graphene\cite{Chengsg,Greenbaum,CXBai,addr1,addr2,Komatsu,Sahu,ChaoWang}, bilayer graphene\cite{Efetov1,Efetov2,Soori}, topological insulator\cite{Majidi}, Weyl semimetal\cite{Chen,Hou,Azizi}, nodal-line semimetal\cite{Cheng}, etc.

The scattering processes of RAR and SAR are distinct, but it is difficult to distinguish them
from the conductance of the metal-SC junction.
For example, the conductance spectra for the three dimensional nodal-line semimetal$-$SC junction possess the same characteristics with those for the graphene-SC junction\cite{Cheng}, although both RAR and SAR are present in the former junction while only RAR or SAR is possible in the latter junction. It has been reported that the shot noise and Fano factor\cite{Zhang} or the Aharonov-Bohm oscillation\cite{Schelter} can be used to characterize the transition between RAR and SAR in the graphene-based SC structures in the presence of ferromagnetic exchange or external field. However, a more universal method of SAR detection irrespective of specific materials is still lack.

The essential difference of RAR and SAR lies in whether the sign relation between the wave vector and the group velocity of the reflected hole is the same with that of the incident electron. We assume the wave vector and the group velocity of the incident electron are of the same (opposite) sign,
then if the two quantities of the reflected hole also have the same (opposite) sign, SAR will happen. On the other hand, if the two quantities of the reflected hole have the opposite (same) sign, RAR will happen. This is because the wave vector component along the junction interface is conserved in the scattering process.

Here, we propose a simple double-junction structure (the metal$-$metal$-$SC junction) to detect SAR without the help of external fields or other interactions, which is completely base on the intrinsic nature of SAR itself. The formation of bound states in the middle metal region of the double junctions is sensitive to the sign relations of the reflected hole. The quasiclassical quantization conditions for energy levels of the bound states are established. The conditions are strongly dependent on the length of the middle metal region for SAR but not for RAR, which lead to the periodic oscillations of conductance when SAR happens. The positions of oscillation peaks and their period can be predicted precisely by the quantization conditions. The establishment of the conditions is irrespective of specific materials and supported by the numerical results. However, to carry out numerical calculations, we construct the double junctions using the recently realized two-dimensional Weyl nodal-line semimetal (WNSM)\cite{Feng,Nie,Niu,Jin} in which the Andreev reflection is of the purely specular type.

The rest of this paper is organized as follows. In Sec. \uppercase\expandafter{\romannumeral2}, the SAR in WNSM$-$SC single junctions with different superconducting pairings and its dependences on the quasiparticle energy, the incident angle and the interfacial barrier are studied. The conductance spectra are presented, which show the same features with those for the RAR. Sec. \uppercase\expandafter{\romannumeral3} builds the quasiclassical quantization conditions for different pairings in the WNSM$-$WNSM$-$SC junctions. The oscillations of SAR and conductance are presented and analysed in details, which have the essential difference from those for the RAR.
Sec. \uppercase\expandafter{\romannumeral4} gives some discussions on our proposed method
and Sec. \uppercase\expandafter{\romannumeral5} concludes this paper.

\section{\label{sec2}SAR and conductance in WNSM$-$SC single junction}
\begin{figure}[!htb]
\centerline{\includegraphics[width=1\columnwidth]{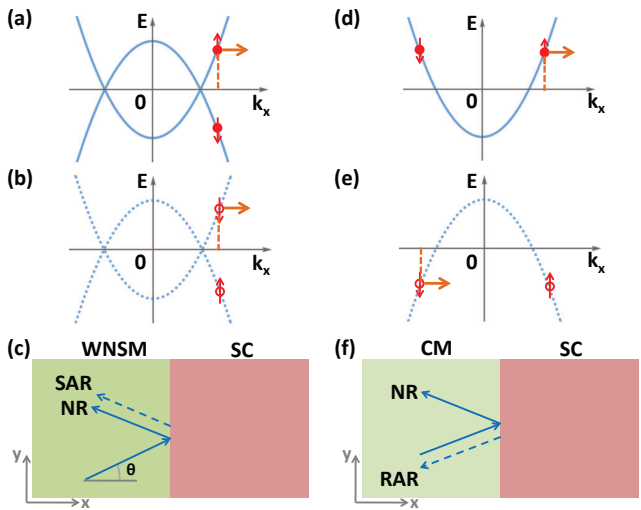}}
\caption{(a) The energy band of WNSM for electrons. The red arrows denote spin. The product of the wave vector and the group velocity (the orange arrow) is positive for the spin-up electron. (b) The energy band of WNSM for holes. The product of the wave vector and the group velocity (the orange arrow) is also positive for the spin-down hole. (c) The scattering processes in the WNSM$-$SC junction. The Andreev reflection is of the purely specular type. (d) The energy band of CM for electrons. The product of the wave vector and the group velocity (the orange arrow) is positive for the spin-up electron. (e) The energy band of CM for holes. The product of the wave vector and the group velocity (the orange arrow) is negative for the spin-down hole. (f) The scattering processes in the CM$-$SC junction. The Andreev reflection is of the purely retro-type.}
\label{fig1}
\end{figure}
We first study the two-dimensional WNSM$-$SC single junction in the $xy$ plane in order to demonstrate the behaviour of the SAR and its equivalent contribution to conductance as the RAR. The junction consists of the semi-infinite WNSM and SC, as well as an interface located at $x=0$ as schematically shown in Fig. \ref{fig1}. The interfacial barrier is expressed as $U(x)=U_{0}\delta(x)$ by the Dirac delta function $\delta(x)$ and the barrier magnitude $U_{0}$.

\subsection{Hamiltonian and wave functions for WNSM}
The Hamiltonian in the spin space of electrons for WNSM can be written as\cite{Jin}
\begin{eqnarray}
\hat{H}_{WNSM}(k)=\left(\begin{array}{cc}
\frac{\hbar^{2} k^2}{2m}-\mu_{W}&0\\
0&-\frac{\hbar^{2} k^2}{2m}+\mu_{W}\\
\end{array}\right).\label{WH}
\end{eqnarray}
Here $m$ is the effective mass, $\mu_{W}$ the material-dependent parameter characterizing the size of the nodal line and ${\bf{k}}=(k_{x},k_{y})$ the wave vector of electrons. Since the Hamiltonian breaks the time-reversal symmetry, the spin degeneracy is lifted. The dispersion for electrons are plotted in Fig. 1(a).

The Bogoliubov-de Gennes(BdG) Hamiltonian in the particle-hole$\otimes$spin space can be written as
\begin{eqnarray}
\check{H}_{WNSM}=\left(\begin{array}{cc}
\hat{H}_{WNSM}(k)&0\\
0&-\hat{H}^*_{WNSM}(-k)\\
\end{array}\right).
\end{eqnarray}
From the BdG Hamiltonian, it is easy to find that the spin-down (up) hole has the same sign relation of the wave vector and the group velocity with the spin-up (down) electron. For example, both wave vector and group velocity of the left-going (right-going) spin-down hole are negative (positive), and their product is positive regardless of the moving direction, as shown in Fig. \ref{fig1}(b). For the spin-up electron, the product of the wave vector and group velocity is also positive [see Fig.\ref{fig1}(a)], which is the same as that of the spin-down hole.
Due to the sign of the product for the spin-up electron and spin-down hole being the same, this determines that the Andreev reflection in WNSM is the SAR as shown in Fig. \ref{fig1}(c).

The sign relation for a CM is totally different. Consider a CM with the BdG Hamiltonian
\begin{eqnarray}
\check{H}_{CM}(k)=\left(\begin{array}{cc}
\hat{H}_{CM}(k)&0\\
0&-\hat{H}_{CM}^{*}(-k)\\
\end{array}\right),\label{CMH}
\end{eqnarray}
in which $\hat{H}_{CM}(k)=(\frac{\hbar^{2} k^2}{2m}-\mu_{N})1_{2\times2}$ with $1_{2\times2}$ the $2\times2$ identity matrix.
The product of the wave vector and group velocity for the spin-up electron is positive [see Fig.\ref{fig1}(d)], but the product for the spin-down hole is negative [Fig.\ref{fig1}(e)]. This sign difference leads to that the Andreev reflection is the RAR
as shown in Fig. \ref{fig1}(f). In fact, the same (opposite) sign relation for the incident electron
and the reflected hole is the intrinsic character of the SAR (RAR), which is independent of the specific materials\cite{Beenakker,Chengsg}.
In this paper, we will propose a method to detect the SAR and RAR based on the same and opposite sign relations.

Through solving the BdG equation $\check{H}_{WNSM}(-i\frac{\partial}{\partial x},k_{y})\psi(x)=E\psi(x)$ with the substitution of $-i\frac{\partial}{\partial x}$ for $k_{x}$, the wave functions for electrons and holes can be obtained. Due to the translation invariance of the junctions along the $y$ axis, the component $k_{y}$ is conserved in the scattering processes.
When a spin-up electron is injected from WNSM to the interface, both spin-down hole and spin-up electron are specularly reflected. The wave function in WNSM with $x<0$ is solved as
\begin{eqnarray}
\psi_{1}=\left(\begin{array}{c}
1\\
0\end{array}\right)e^{ik^{e\uparrow}_{x}x}
+a_{\downarrow}\left(\begin{array}{c}
0\\
1\end{array}\right)e^{-ik^{h\downarrow}_{x}x}
+b_{\uparrow}\left(\begin{array}{c}
1\\
0\end{array}\right)e^{-ik^{e\uparrow}_{x}x},\label{wfwnsm}
\end{eqnarray}
with the SAR coefficient $a_{\downarrow}$ and the normal reflection coefficient $b_{\uparrow}$. The wave vectors are given by $k_{x}^{e\uparrow}(k_{x}^{h\downarrow})=\sqrt{\frac{2m}{\hbar^2}(\mu_{W}+ E)-k_{y}^2}$ with $k_{y}=\sqrt{\frac{2m}{\hbar^2}(\mu_{W}+E)}\sin\theta$. The angle $\theta$ is the incident angle of the electron as shown in Fig. 1(c). The wave function $\psi_{2}$ for the spin-down electron incidence can be obtained in a similar way. We use $a_{\uparrow}$ and $b_{\downarrow}$ to denote SAR and the normal reflection in this scattering process, respectively.

\subsection{Hamiltonian and wave functions for SC}
For SC, we consider the $s$-wave, $d$-wave and chiral $p$-wave pairings. The BdG Hamiltonian for the SCs is given by
\begin{eqnarray}
H_{SC}=\left(\begin{array}{cccc}
\hat{\epsilon}({\bf{k}})&\hat{\Delta}({\bf{k}})\\
-\hat{\Delta}^{*}(-{\bf{k}})&-\hat{\epsilon}({\bf{-k}}),
\end{array}\right),
\end{eqnarray}
where the single-particle energy $\hat{\epsilon}({\bf{k}})=(\frac{\hbar^2 k^2}{2m}-\mu_{S})\hat{1}_{2\times2}$, the energy matrix $\hat{\Delta}({\bf{k}})=\Delta_{0}f({\bf{k}})i\hat{\sigma}_{y}$ for the $s$-wave or $d$-wave SC and $\hat{\Delta}({\bf{k}})=\Delta_{0}f({\bf{k}})\hat{\sigma}_{x}$ for the $p$-wave SC. Here, $\hat{\sigma}_{x}$ and $\hat{\sigma}_{y}$ are Pauli matrices in the spin space. For the $s$, $d_{x^2-y^2}$ and $d_{xy}$-wave pairings\cite{KashiwayaRPP,Hirai}, $f({\bf{k}})=1$, $\hat{k}_{x}^2-\hat{k}_{y}^2$ and $2\hat{k}_{x}\hat{k}_{y}$, respectively, while for the $p$-wave pairing\cite{Hirai,Mackenzie}, $f({\bf{k}})=\hat{k}_{x}+i\hat{k}_{y}$. The effective mass for SCs has been taken as the same with that for WNSM and $\mu_{S}$ is the chemical potential.

Through solving the BdG equation, we can obtain the wave functions in the SC region with $x>0$. For the spin-up electron incidence, the function is written as
\begin{eqnarray}
\Psi_{1}=c_{\uparrow}\left(\begin{array}{c}
u_{+}e^{i\phi_{+}}\\
v_{+}\end{array}\right)e^{ik_{x}x}
+d_{\downarrow}\left(\begin{array}{c}
v_{-}e^{i\phi_{-}}\\
u_{-}\end{array}\right)e^{-ik_{x}x}.
\end{eqnarray}
Here, $c_{\uparrow}$ and $d_{\downarrow}$ represent the transmissions as electron-like quasiparticle and hole-like quasiparticle, respectively. The wave vector is $k_{x}=\sqrt{\frac{2m}{\hbar^2}\mu_{S}-k_{y}^2}$
under the Andreev approximation\cite{Andreev}. For the $s$-wave and $p$-wave SCs, $u_{+/-}=\sqrt{\frac{E+\Omega}{2E}}$ and $v_{+/-}=\sqrt{\frac{E-\Omega}{2E}}$ with $\Omega=\sqrt{E^2-\Delta_{0}^2}$. For the $d$-wave SC, $u_{\pm}=\sqrt{\frac{E+\Omega_{\pm}}{2E}}$ and $v_{\pm}=\sqrt{\frac{E-\Omega_{\pm}}{2E}}$ with $\Omega_{\pm}=\sqrt{E^2-\Delta_{\pm}^2}$. The direction dependent energy gap $\Delta_{\pm}=\Delta_{0}\cos{[2(\theta_{s}\mp\beta)]}$ in which $\theta_{s}$ is the transmission angle for electron-like quasiparticles and $\beta$ is the angle between the interface normal and the crystallographic $a$-axis of the $d$-wave SC. The angle $\theta_{s}$ can be expressed by $\theta$ under the conservation of $k_{y}$. For the $d_{x^2-y^2}$-wave SC, $\beta=0$ and for the $d_{xy}$-wave SC, $\beta=\frac{\pi}{4}$. The internal phase factors $e^{i\phi_{\pm}}$ are $1$, $\frac{\cos(2\theta_{s}\mp2\beta)}{|\cos(2\theta_{s}\mp2\beta)|}$ and $\frac{\pm\cos{\theta_{s}}+ i\sin{\theta_s}}{|\cos{\theta_s}+i\sin{\theta_s}|}$ for the $s$-wave pairing, the $d$-wave pairing and the $p$-wave pairing, repsectively. For the spin-down electron incidence, the wave function $\Psi_2$ for SC can be derived in a similar way.

\subsection{Boundary conditions and conductance}
The boundary conditions which ensure the probability conservation for the WNSM-SC single junction are given by
\begin{eqnarray}
\begin{split}
&\psi_{1(2)}|_{x=0^-}=\Psi_{1(2)}|_{x=0^+},\\
&\Psi'_{1}|_{x=0^+}-\hat{\tau}_{z}\psi'_{1}|_{x=0^-}=\frac{2mU}{\hbar^2}\psi_{1}|_{x=0},\\
&\Psi'_{2}|_{x=0^+}+\hat{\tau}_{z}\psi'_{2}|_{x=0^-}=\frac{2mU}{\hbar^2}\psi_{2}|_{x=0},\label{WSBC}
\end{split}
\end{eqnarray}
with $\hat{\tau}_{z}$ being the Pauli matrix in the particle-hole space. Under the boundary conditions, the reflection coefficients $a_{\uparrow}$, $a_{\downarrow}$, $b_{\uparrow}$ and $b_{\downarrow}$ can be solved (see Appendix).

We define the probabilities for SAR and the normal reflection as
\begin{eqnarray}
A_{\uparrow}=\text{Re}\left[\frac{k_{x}^{h\uparrow}}{k_{x}^{e\downarrow}}\right]|a_{\uparrow}|^2,~~~B_{\uparrow}=|b_{\uparrow}|^2,\\
A_{\downarrow}=\text{Re}\left[\frac{k_{x}^{h\downarrow}}{k_{x}^{e\uparrow}}\right]|a_{\downarrow}|^2,~~~B_{\downarrow}=|b_{\downarrow}|^2.
\label{prob}
\end{eqnarray}
According to the Blonder-Tinkham-Klapwijk theory\cite{Blonder}, the conductance can be expressed as
\begin{eqnarray}
\sigma_{\uparrow}=1+A_{\downarrow}-B_{\uparrow},\\
\sigma_{\downarrow}=1+A_{\uparrow}-B_{\downarrow},\label{fencond}
\end{eqnarray}
which are caused by the spin-up electron incidence and the spin-down electron incidence, respectively.
The total conductance normalized by the normal conductance is given by
\begin{eqnarray}
\sigma(eV)=\frac{\int_{-\pi/2}^{\pi/2}(\sigma_{\uparrow}+\sigma_{\downarrow})\cos{\theta}d\theta}
{\int_{-\pi/2}^{\pi/2}(\sigma_{n\uparrow}+\sigma_{n\downarrow})\cos{\theta}d\theta},\label{totalcond}
\end{eqnarray}
where $\sigma_{n\uparrow}$ and $\sigma_{n\downarrow}$ are the conductance when SCs are in the normal state and $V$ is the bias of the junction. Note, the zero temperature conductance is considered in this paper. In this situation, we have the relations $E=eV$ and $\sigma(E)=\sigma(eV)$ in accordance with the Blonder-Tinkham-Klapwijk theory\cite{Blonder}.

\subsection{\label{sec3}Numerical results}
For simplicity, we set $\mu_{W}=\mu_S=\mu$ in our calculations. The mismatch between $\mu_{W}$ and $\mu_{S}$ will suppress the conductance but will not fundamentally change our physical results. Since the conductance are independent on the spin of the incident electron, we only show the numerical results of probability $A_{\downarrow}$ for the spin-up electron incidence. We define the effective interfacial barrier as $z=\frac{2mU_{0}}{\hbar^2k_{F}}$ with $k_{F}=\sqrt{\frac{2m\mu}{\hbar^2}}$.

\begin{figure}[!htb]
\centerline{\includegraphics[width=0.9\columnwidth]{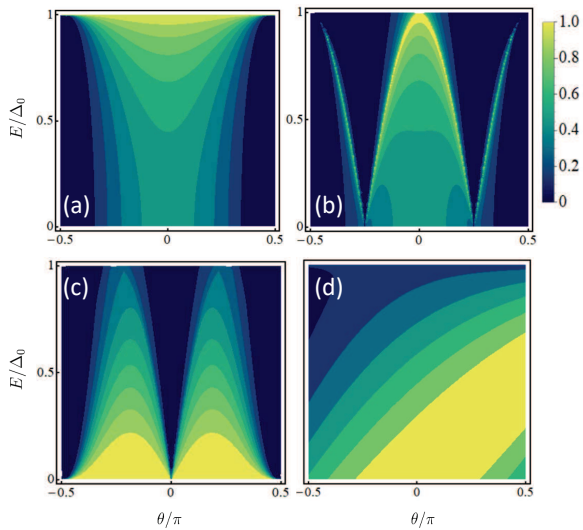}}
\caption{The incident angle $\theta$ and the incident energy $E$ dependences of the SAR probability at $z=1$ for (a) the $s$-wave pairing, (b) the $d_{x^2-y^2}$-wave pairing, (c) the $d_{xy}$-wave pairing and (d) the $p$-wave pairing.}
\label{SAR}
\end{figure}

Now, we discuss the incident angle $\theta$ and the quasiparticle energy $E$ dependences of SAR. For the transparent junction with $z=0$, the subgap SAR probability $A_{\downarrow}$ for the $s$-wave and the $p$-wave SCs is always $1$, which is irrespective of $\theta$ and $E$. This will lead to the normalized conductance of the value $2$ in the whole gap. However, for the $d$-wave SCs, the $\theta$-independent SAR with $A_{\downarrow}=1$ only happens at $E=0$. As a result, the zero-bias conductance (ZBC) is still $2$ but the conductance for $eV>0$ is suppressed.

\begin{figure}[!htb]
\centerline{\includegraphics[width=0.9\columnwidth]{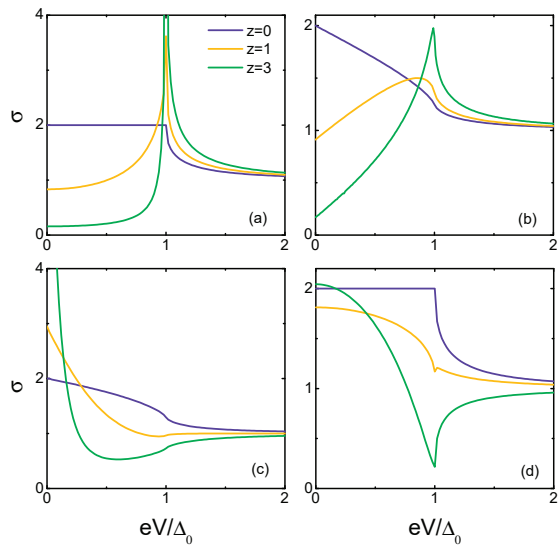}}
\caption{The normalized conductance spectra at different interfacial barriers for (a) the $s$-wave pairing, (b) the $d_{x^2-y^2}$-wave pairing, (c) the $d_{xy}$-wave pairing and (d) the $p$-wave pairing.}
\label{cond}
\end{figure}

For the nonzero interfacial barrier with $z\ne0$, we present the SAR probabilities for $z=1$ in Fig. {\ref{SAR}}. For the $s$-wave pairing in Fig. \ref{SAR}(a), it is found that the probability $A_{\downarrow}$ is dramatically weakened for the small energy near $E=0$ compared with the SAR of $z=0$. When the incident energy $E$ is increased from 0 to $\Delta_{0}$, the SAR probability tends to $1$. This means the conductance increases with the bias in the gap. The similar thing happens for the $d_{x^2-y^2}$-wave pairing in Fig. {\ref{SAR}}(b). The conductance spectra at $z=0$ and $z=1$ for the $s$-wave and the $d_{x^2-y^2}$-wave pairings are shown in Figs. {\ref{cond}}(a) and (b). The behaviours of the conductance are consistent with the above analyses on SAR.
The situations for the $d_{xy}$-wave and the $p$-wave pairings become very different. The SARs are reduced near $E=\Delta_{0}$ when one increases $z$ from $0$ to $1$, but they keep the value $A_{\downarrow}\sim 1$ in a large range of the incident angle near $E=0$ as shown in Figs. \ref{SAR}(c) and (d). The conductance will decrease as the bias is increased in the gap. The conductance spectra for $z=0$ and $z=1$ are shown in Figs. \ref{cond}(c) and (d).

If the interfacial barrier continues to increase, for example to $z=3$, the features of SAR presented in Fig. \ref{SAR} will become more obvious. The resulting conductance spectra are plotted in Figs. \ref{cond}(a)-(d) for different pairings. For the $s$-wave and the $d_{x^2-y^2}$-wave pairings, the shapes of conductance evolve towards the bulk density of states of SCs\cite{Blonder,Tanaka}. For the $d_{xy}$-wave and the $p$-wave pairings, the conductance possesses the ZBC peaks, which characterize the existence of the bound states at the surface of SCs\cite{Kashiwaya}. It can be concluded that the SAR can also well reflect the anisotropic properties of SCs as RAR.

The SAR in the WNSM$-$SC single junction possesses the same features with the RAR in the CM$-$SC single junction. Beneficially, we can obtain the same amplitude of SAR in WNSM as RAR in CM. Unfortunately, as an observable quantity, the conductance for the WNSM$-$SC junction is of the same form with that for the CM$-$SC junction\cite{Blonder,Tanaka,Yamashiro}. This indicates the contributions of the SAR and RAR to the conductance are equivalent. It is difficult to distinguish the two types of reflections by measuring the electric transport of the single-junction structure.

\section{\label{sec4}Detection of SAR in WNSM$-$WNSM$-$SC junctions}

\begin{figure}[!htb]
\centerline{\includegraphics[width=0.7\columnwidth]{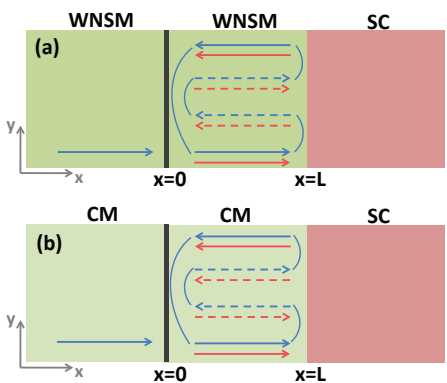}}
\caption{(a) Schematic illustration of the WNSM-WNSM-SC double junctions and the scattering processes in the middle WNSM for the electron incidence from the left WNSM. The group velocities are denoted by blue lines which form a closed path. Along the path, the wave vectors are denoted by red lines, correspondingly. (b) The same closed path (blue lines) and the corresponding wave vectors (red lines) along the path in the CM-CM-SC double junctions. In the figure, the solid lines for electron and the dashed ones for hole.
Here the directions of the wave vectors for holes in (a) and (b) are
opposite.}
\label{wws}
\end{figure}

In order to distinguish SAR and RAR, we consider the WNSM$-$WNSM$-$SC double-junction structure as shown in Fig. \ref{wws}. The two interfaces are located at $x=0$ and $x=L$, which potential can be described by $U(x)=U_{1}\delta(x)+U_{2}\delta(x-L)$. When an electron is injected from the left WNSM, it will transmit into the middle WNSM through the interface at $x=0$. The transmitted electron impings on the interface at $x=L$ and is specularly reflected as the left-going hole. The left-going hole is normally reflected as the right-going one at the interface with $x=0$ and then specularly reflected as the left-going electron at the interface with $x=L$. The left-going electron will be normally reflected as the right-going one at the interface $x=0$. After twice SARs and twice normal reflections, a closed path of the quasiparticle motion is formed as depicted in Fig. \ref{wws}(a).

\subsection{$z_{2}=0$}
We first consider the junctions with $z_{2}=0$ to clarify the basic physics.
For the $s$-wave and $d_{x^2-y^2}$-wave SCs, if the phase accumulated along the closed path satisfies the following quasiclassical quantization condition,
\begin{eqnarray}
-\sum_{\alpha=\pm}\arccos{\frac{E}{|\Delta_{\alpha}|}}+2(k_{x}^{e\uparrow}+k_{x}^{h\downarrow})L=2n\pi,\label{sdqqc}
\end{eqnarray}
with $n$ an integer number, the energy levels of bound states will be formed. For the $s$-wave pairing, $\Delta_{\pm}=\Delta_{0}$ while for the $d_{x^2-y^2}$-wave pairing, $\Delta_{\pm}=\Delta_{0}\cos(2\theta\mp2\beta)$ with $\beta=0$. The wave vectors $k_{x}^{e\uparrow}$ and $k_{x}^{h\downarrow}$ are the ones in Eq. (\ref{wfwnsm}). The first term in Eq. (\ref{sdqqc}) is acquired from the Andreev reflection (see Appendix for details). The second term originates from the motion of quasiparticles along the closed path. For the zero energy bound state with $E=0$ and the normal incidence with $k_{y}=0$, the quantization condition degenerates into
\begin{eqnarray}
k_{F}L=\frac{(2n+1)\pi}{4},\label{p1}
\end{eqnarray}
with $k_{F}=\sqrt{\frac{2m\mu}{\hbar^2}}$. This indicates ZBC of the junctions will oscillate with the length $L$ of the middle WNSM. The peaks of the ZBC appear at $k_{F}L=\frac{\pi}{4}, \frac{3\pi}{4}, \frac{5\pi}{4}\cdots.$ The oscillation period defined as the spacing between the neighbouring peaks is $\Delta k_{F}L=\frac{\pi}{2}$.

For the $d_{xy}$-wave and the $p$-wave SCs, the quasiclassical quantization condition becomes
\begin{eqnarray}
\pi-\left[\sum_{\alpha=\pm}\arccos{\frac{E}{|\Delta_{\alpha}|}}+2\delta\theta\right]
+2(k_{x}^{e\uparrow}+k_{x}^{h\downarrow})L=2n\pi,\label{pdqqc}
\end{eqnarray}
where $\Delta_{\pm}=\Delta_{0}$ and $\delta=1$ for the $p$-wave SC while $\Delta_{\pm}=\Delta_{0}\cos[2(\theta\mp\beta)]$ with $\beta=\frac{\pi}{4}$ and $\delta=0$ for the $d_{xy}$-wave SC. The difference between Eq. (\ref{sdqqc}) and Eq. (\ref{pdqqc}) is caused by the anisotropy of the $p$-wave and the $d$-wave SCs (see Appendix for details). The phase $(-\arccos{\frac{E}{|\Delta_{+}|}}-\delta\theta)$ is acquired by the SAR with the conversion from electron to hole while the phase $(\pi-\arccos{\frac{E}{|\Delta_{-}|}}-\delta\theta)$ is acquired by the SAR with the conversion from hole to electron as shown in Fig. {\ref{wws}}(a).
For the $s$-wave and the $d_{x^2-y^2}$-wave SCs, the two phases are the same. According to Eq. (\ref{pdqqc}), the zero energy levels for the normal incidence ($\theta=0$) can be formed when
\begin{eqnarray}
k_{F}L=\frac{2n\pi}{4},\label{p2}
\end{eqnarray}
which indicates the peaks of the ZBC appear at $k_{F}L=0,\frac{\pi}{2},\pi\cdots$. The oscillation period is $\Delta k_{F}L=\frac{\pi}{2}$.

The expressions in Eqs. (\ref{sdqqc}) and (\ref{pdqqc}) are our critical results which can be used to distinguish SAR from RAR. The key point is the presence of the term $2(k_{x}^{e\uparrow}+k_{x}^{h\downarrow})L$ derived from the integration of wave vectors along the closed path. This term embodies the intrinsic character of the specularly Andreev reflected hole. For the spin-up  electron in the middle WNSM, its wave vectors and its group velocities have the same sign, i.e., they are in the same direction, see Figs. \ref{fig1}(a) and \ref{wws}(a).
The specularly reflected spin-down hole also possesses the relation. Its wave vectors and group velocities are also in the same direction as shown in
Figs. \ref{fig1}(b) and \ref{wws}(a).
As a result, along the closed path, the phase acquired by the motion of the hole is $2k_{x}^{h\downarrow}L$ when the phase for the electron is $2k_{x}^{e\uparrow}L$.

However, it is not the case for RAR in CM-CM-SC double junctions since the retro-Andreev reflected hole has the opposite sign relation with the incident electron as shown in Figs. \ref{fig1}(d), \ref{fig1}(e) and \ref{wws}(b).
For the spin-up electron in the middle CM, its wave vectors and its group velocities have the same sign, but the wave vectors and group velocities of the retro-reflected hole have the opposite sign\cite{Zagoskin,Chengphyc,ChengJLTP}.
Along the closed path, the phase acquired by the motion of the hole is $-2k_{x}^{h\downarrow}L$ when the phase for the electron is $2k_{x}^{e\uparrow}L$. For $E=0$, we have $k_{x}^{e\uparrow}=k_{x}^{h\downarrow}$ exactly. The phase from the quasiparticle motion is zero. The quantization condition is independent on the length of CM. There is no oscillation effect for ZBC. Even if $E\ne0$, the conclusion still holds as long as $E$ is much smaller than the chemical potential.

\begin{figure}[!htb]
\centerline{\includegraphics[width=0.9\columnwidth]{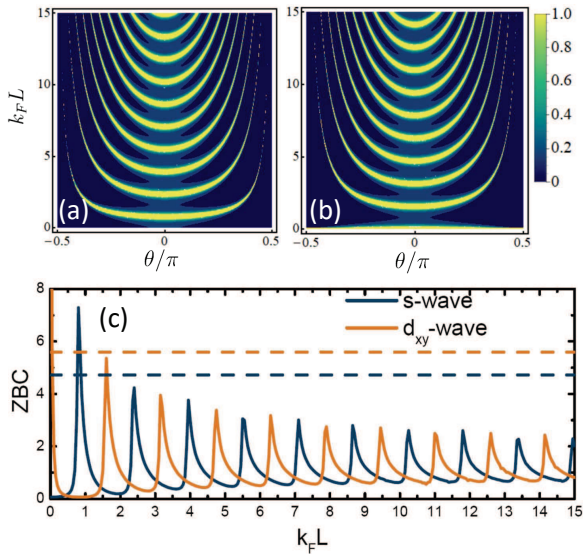}}
\caption{The length $k_{F}L$ and the incident angle $\theta$ dependences of SAR at $E=0$ for (a) the $s$-wave pairing and (b) the $d_{xy}$-wave pairing. (c) The oscillation of ZBC as the length $k_{F}L$ for the $s$-wave pairing and the $d_{xy}$-wave pairing in the WNSM-WNSM-SC junctions. The effective interfacial barriers have been taken as $z_{1}=5$ and $z_{2}=0$. The width-independent straight dashed lines denote the ZBCs for the $s$-wave pairing (the lower one) and the $d_{xy}$-wave pairing (the upper one) in the CM-CM-SC junctions under the same parameters.}
\label{osi}
\end{figure}

To support our physical analyses, we calculate numerically SAR and conductance in WNSM-WNSM-SC double junctions. We define the effective interfacial barriers as $z_{1}=\frac{2mU_{1}}{\hbar^2k_{F}}$ and $z_{2}=\frac{2mU_{2}}{\hbar^2k_{F}}$. In the calculations, $z_{1}=5$ and $z_{2}=0$ are taken. Fig. {\ref{osi}}(a) and (b) show the length and the incident angle dependences of SAR at $E=0$ for the $s$-wave pairing and the $d_{xy}$-wave pairing, respectively. For $k_{F}L \rightarrow 0$, SAR is almost completely suppressed for the $s$-wave pairing while it can happen for the $d_{xy}$-pairing. The former corresponds to the ZBC near to zero and the latter corresponds to the ZBC peak. They are consistent with the results of the WNSM-SC single junction as given in Figs. \ref{cond}(a) and (c).

The oscillation of the SAR probabilities with the length of the middle WNSM is very obvious for a given incident angle $\theta$, which can be well explained from the quantization conditions Eqs. (\ref{sdqqc}) and (\ref{pdqqc}). The bright fringes demonstrate the formation of the zero energy bound states. Each bright fringe corresponds to an integer number $n$. From the bottom to the top, $n$ is $0,1,2\cdots$. For a certain bright fringe with the number $n$, it bends in the direction of the length increase as the incident angle is raised. This is because the wave vectors $k_{x}^{e\uparrow}$ and $k_{x}^{h\downarrow}$ are reduced when one raises the incident angle $\theta$. According to Eqs. (\ref{sdqqc}) and (\ref{pdqqc}), we need to enlarge the length $L$ to ensure the establishment of the equations with the fixed number $n$.

Fig. \ref{osi}(c) presents the conductance integrated about $\theta$ for the $s$-wave pairing and the $d_{xy}$-wave pairing. Although the integration about $\theta$ will bring about broadening of the ZBC peaks, the positions of peaks still fit Eqs. (\ref{p1}) and (\ref{p2}) very well. The $\frac{\pi}{4}$ shift of the peaks for the $d_{xy}$-pairing relative to those for the $s$-wave pairing is derived from the extra $\pi$ phase in Eq. (\ref{pdqqc}) due to the anisotropy of the $d$-wave SC. This shift also shows itself in SAR as given in Fig. \ref{osi}(a) and (b). The SAR and ZBC for the $d_{x^2-y^2}$-wave and the $p$-wave pairings are not presented here since the former is similar to the $s$-wave case and the latter is similar to the $d_{xy}$-wave case. In addition, the measurement on the period of the discrete ZBC peaks can provide the immediate information about the nodal-line size in WNSM.

For comparision, we also present the ZBCs of the CM-CM-SC junctions for the $s$-wave pairing (the lower straight dashed line) and the $d_{xy}$-wave pairing (the upper straight dashed line) in Fig. \ref{osi}(c). They are both independent on the width of the middle CM as we have discussed. Actually, the width-independent RAR for the zero energy in the CM-CM-SC junctions are heavily suppressed for the $s$-wave pairing by the high barrier with $Z_{1}=5$, while the probability of RAR is still $1$ for the $d_{xy}$-wave pairing. Accordingly, the ZBC for the $s$-wave pairing is almost zero and that for the $d_{xy}$-pairing has a peak with a large value. In Fig. \ref{osi}(c), the value for the lower dashed line has been magnified eighty times when the value for the upper dashed line has been reduced five times. These behaviors of RAR and ZBC for the CM-CM-SC double junctions are the same as those for the CM-SC single junction but totally different from the features for the WNSM-WNSM-SC double junctions.

\subsection{$z_{2}\ne0$}
\begin{figure}[!htb]
\centerline{\includegraphics[width=0.9\columnwidth]{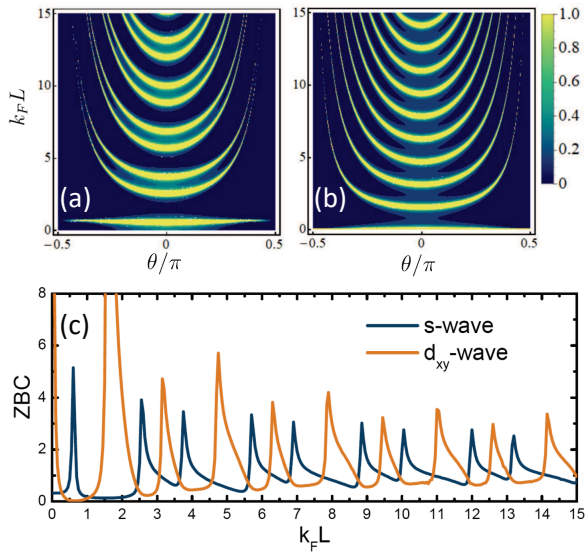}}
\caption{The length $k_{F}L$ and the incident angle $\theta$ dependences of SAR at $E=0$ for (a) the $s$-wave pairing and (b) the $d_{xy}$-wave pairing. (c) The oscillation of ZBC as the length $k_{F}L$ for the $s$-wave pairing and the $d_{xy}$-wave pairing. The effective interfacial barriers have been taken as $z_{1}=5$ and $z_{2}=2$.}
\label{osi2}
\end{figure}
Finally, we consider the more realistic junctions with $z_{2}\ne0$. Fig. \ref{osi2} shows SAR probabilities and ZBC for $z_{1}=5$ and $z_{2}=2$ at $E=0$. For the $d_{xy}$-wave pairing, both SAR and ZBC oscillations remain virtually unchanged, which include the positions of peaks and the period. However, there are significant changes for the $s$-wave pairing. First, the ZBC peaks are greatly weakened, which have been magnified ten times in Fig. \ref{osi2}(c). Second, the distribution of peaks is no longer uniform. The spacing between the $2n$th and the $(2n+1)$th peaks with $n\geq1$ becomes small. This can also be seen from the SAR probability in Fig. \ref{osi2}(a). Actually, the spacing will get smaller and smaller as $z_{2}$ increases. Two neighbor peaks will form a main peak with split when $z_{2}$ is large enough, for example $z_{2}=5$. The two changes result from the suppression of SAR and the enhancement of normal reflection of the $s$-wave pairing when the effective barrier $z_{2}$ is raised at $E=0$.

The normal reflection in the middle WNSM also leads to conductance oscillation. Twice normal reflections, one at $x=0$ and the other at $x=L$, will form a closed path. The quasiclassical quantization condition is simply $k_{F}L=n\pi$ for the normal incidence with $\theta=0$. Hence, the positions of main peaks are located at $k_{F}L=0,\pi,2\pi,\cdots$. Each main peak possesses two split peaks from SAR. The similar thing also happens for the $d_{x^2-y^2}$ wave and the $p$-wave pairings. It is just that the peaks for the $p$-wave pairing are not weakened so much since there is still a range of incident angle for strong SAR at $E=0$ as given in Fig. \ref{SAR}(d). The situation is different for the $d_{xy}$-wave pairing which ZBC peaks from SAR can survive even if $z_{2}=5$. The total SAR at $E=0$ is unaffected by the interfacial barrier as shown in Fig. \ref{SAR}(c). In brief, the SAR induced oscillation is completely distinguishable for all pairings at least when $z_{2}\leq2$. For more large $z_{2}$, the split of main peaks is still a strong signal for SAR.

\section{\label{secdiscussion} Discussions}

Let us give short discussions on the detection of SAR in other materials.
For WNSM here, we study the behaviors of ZBC, which are related to the formation of zero energy Andreev levels. In this situation, the conductance oscillates for SAR while it does not for RAR.
However, the zero energy is not the necessary requirement to distinguish the two Andreev reflections.
Even if the finite bias conductances with $E\ne0$ are considered,
which will oscillate both for RAR and SAR, their oscillation behaviors are still distinct.
The key is that the quantization conditions for the levels of Andreev bound states are different for RAR and SAR due to their intrinsic characters.
The conditions will form different Andreev levels for RAR and SAR in the junction parameter space,
which can show themselves in conductances and be detected.
For materials such as the monolayer graphene\cite{Beenakker}, bilayer graphene\cite{Efetov2}, topological insulator\cite{Majidi} and nodal-line semimetal\cite{Cheng}, the establishment of quantization conditions could be a little more complicated, but our method is still effective,
which can still predict some distinguishable features for RAR and SAR.

Finally, we discuss the difference between the CM$-$CM$-$SC (WNSM$-$WNSM$-$SC) junctions
and the SC$-$CM$-$SC (SC$-$WNSM$-$SC) junctions
to show that the proposed method to distinguish the RAR and SAR
can only work in the WNSM$-$WNSM$-$SC junctions.
Actually, for clean SC$-$CM(WNSM)$-$SC junctions with transparent interfaces,
the quantization condition for the Andreev levels can be simply written as
(here we take the $s$-wave SC as an example)\cite{Zagoskin}
\begin{eqnarray}
-2\arccos{\frac{E_{n}^{\pm}}{\Delta_{0}}}\pm\varphi
+(k_{x}^{e\uparrow}-k_{x}^{h\downarrow})L=2n\pi ,\label{R1}
\end{eqnarray}
for the RAR with the superconducting phase difference $\varphi$,
and the quantization condition is
\begin{eqnarray}
-2\arccos{\frac{E_{n}^{\pm}}{\Delta_{0}}}\pm\varphi
+(k_{x}^{e\uparrow}+k_{x}^{h\downarrow})L=2n\pi ,\label{R2}
\end{eqnarray}
for the SAR.
Although the quantization conditions in Eqs.(\ref{R1}) adn (\ref{R2}) are similar with
that in Eq.(\ref{sdqqc}),
this dose not mean that the Josephson current from the RAR in the SC$-$CM(WNSM)$-$SC junctions
is not dependent on the length $L$ because the Josephson current includes contributions from all Andreev bound states with $E_{n}^{\pm}<0$
even at zero temperature.
From Eq.(\ref{R1}), the formation of the zero energy level $E=0$ is irrespective of the length $L$,
but the levels with $E\ne 0$ is dependent on the length $L$, leading to the Josephson current
depending on the length $L$ also.

\section{\label{sec5}Conclusions}
We propose a more universal method to detect SAR by using
the semimetal-semimetal-SC double-junction structure.
The method is completely base on the intrinsic character of
the specularly reflected hole in SAR which has the same sign relation
about the group velocity and the wave vector with the incident electron.
However, the sign relation for the retro-Andreev reflected hole in RAR
is opposite. This essential difference between SAR and RAR
can be seized subtly by the accumulated phase of the quasipartilce
motion along a close path in the middle semimetal.
Through establishing the quasiclassical quantization conditions
for the energy levels of bound states, one finds the ZBC from SAR
periodically oscillates with the length of the middle region
while that from RAR does not.
Further, the positions and the period of ZBC peaks can be predicted
precisely with the quantization conditions.
This provides strong distinguishable signatures of SAR.
Actually, even though the phases acquired by electron and hole
do not cancel for RAR at the non-zero incident energy,
the oscillation features of ZBC, including positions and period of peaks,
are still recognizably different from those for SAR.

\section*{\label{sec5}ACKNOWLEDGMENTS}
This work was financially supported by National Key R and D Program of China (2017YFA0303301),
NSF-China under Grants Nos. 11921005 and 11447175,
the Strategic Priority Research Program of Chinese Academy of Sciences (XDB28000000)
and the Natural Science Foundation of Shandong Province under Grants No. ZR2017QA009.

\section{Appendix}
\setcounter{equation}{0}
\setcounter{subsection}{0}
\renewcommand{\theequation}{A.\arabic{equation}}
\renewcommand{\thesubsection}{A.\arabic{subsection}}
\subsection{Derivation of the quasiclassical quantization conditions}
Using the boundary conditions in Eq.(\ref{WSBC}), the SAR and normal reflecton coefficients for the spin-up electron incidence in WNSM-SC junction are solved as
\begin{eqnarray}
\begin{split}
a_{\downarrow}=\frac{4k_{x}k_{x}^{e\uparrow}u_{-}v_{+}}{k_{x}^2\xi_{-}+k_{x}(k_{x}^{e\uparrow}+k_{x}^{h\downarrow})\xi_{+}
+(z-i k_{x}^{e\uparrow})(z+i k_{x}^{h\downarrow})\xi_{-}},\\
b_{\uparrow}=\frac{k_{x}(k_{x}^{e\uparrow}+k_{x}^{h\downarrow})\xi_{+}
-k_{x}^2\xi_{-}+( k_{x}^{e\uparrow}-iz)(k_{x}^{h\downarrow}-iz)\xi_{-}}{k_{x}^2\xi_{-}+k_{x}(k_{x}^{e\uparrow}+k_{x}^{h\downarrow})\xi_{+}
+(z-i k_{x}^{e\uparrow})(z+i k_{x}^{h\downarrow})\xi_{-}},
\end{split}
\end{eqnarray}
with $\xi_{\pm}=u_{+}u_{-}e^{i\phi_{+}}\pm v_{+}v_{-}e^{i\phi_{-}}$.

We consider the transparent junction with $z=0$.
Under the Andreev approximation, we take $k_{x}=k_{x}^{e\uparrow}=k_{x}^{h\downarrow}$. The SAR coefficient degenerates into
\begin{eqnarray}
a_{\downarrow}=\frac{v_{+}}{u_{+}}e^{-i\phi_{+}}.
\end{eqnarray}
For the $s$-wave and the $d_{x^2-y^2}$-wave pairings, $a_{\downarrow}=e^{-i\arccos{\frac{E}{|\Delta_{+}|}}}$ with $E<|\Delta_{+}|$. The expression implies a phase of $(-\arccos{\frac{E}{|\Delta_{+}|}})$ is acquired by the specularly reflected hole in the SAR process.
For the $d_{xy}$-wave and the $p$-wave pairings, the coefficient $a_{\downarrow}=e^{-i\arccos{\frac{E}{|\Delta_{+}|}}}e^{-i\delta\theta}$. This implies the specularly reflected hole will acquire a phase of $(-\arccos{\frac{E}{|\Delta_{+}|}}-\delta\theta)$. Here, $\Delta_{+}=\Delta_{0}$ for the $s$-wave and $p$-wave pairings while $\Delta_{+}=\Delta_{0}\cos(2\theta-2\beta)$ for the $d_{x^2-y^2}$-wave pairing $(\beta=0)$ and the $d_{xy}$-wave pairing $(\beta=\frac{\pi}{4})$. In addition, $\delta=1$ for the $p$-wave SC and $\delta=0$ for the $d_{xy}$-wave SC.

On the other hand, when a spin-down hole is injected from WNSM, the coefficients $\tilde{a}_{\uparrow}$ for the specularly reflected electron and $\tilde{b}_{\downarrow}$ for the normally reflected hole can also be solved as
\begin{eqnarray}
\begin{split}
\tilde{a}_{\uparrow}&=\frac{4k_{x}k_{x}^{h\downarrow}u_{+}v_{-}e^{i\phi_{+}e^{i\phi_{-}}}}
{k_{x}^2\xi_{-}+k_{x}(k_{x}^{e\uparrow}+k_{x}^{h\downarrow})\xi_{+}
+(z-i k_{x}^{e\uparrow})(z+i k_{x}^{h\downarrow})\xi_{-}},\\
\tilde{b}_{\downarrow}&=-\frac{k_{x}(k_{x}^{e\uparrow}-k_{x}^{h\downarrow})\xi_{+}
+k_{x}^2\xi_{-}-(k_{x}^{e\uparrow}+iz)(k_{x}^{h\downarrow}+iz)\xi_{-}}
{k_{x}^2\xi_{-}+k_{x}(k_{x}^{e\uparrow}+k_{x}^{h\downarrow})\xi_{+}
+(z-i k_{x}^{e\uparrow})(z+i k_{x}^{h\downarrow})\xi_{-}}.
\end{split}
\end{eqnarray}
Under the Andreev approximation, $\tilde{a}_{\uparrow}$ becomes
\begin{eqnarray}
\tilde{a}_{\uparrow}=\frac{v_{-}}{u_{-}}e^{i\phi_{-}}.
\end{eqnarray}
For the $s$-wave and $d_{x^2-y^2}$-wave pairings, $\tilde{a}_{\uparrow}=e^{-i\arccos{\frac{E}{|\Delta_{-}|}}}$ with $E<|\Delta_{-}|$, which means a phase of $(-\arccos{\frac{E}{|\Delta_{-}|}})$ is obtained by the specularly reflected electron. For the $d_{xy}$-wave and $p$-wave pairings, $\tilde{a}_{\uparrow}=-e^{-i\arccos{\frac{E}{|\Delta_{-}|}}}e^{-i\delta\theta}$ which means a phase of $(\pi-\arccos{\frac{E}{\Delta_{0}}}-\delta\theta)$ is obtained by the specularly reflected electron. Here, $\Delta_{-}=\Delta_{0}$ for the $s$-wave and $p$-wave pairings while $\Delta_{-}=\Delta_{0}\cos(2\theta+2\beta)$ for the $d_{x^2-y^2}$-wave pairing $(\beta=0)$ and the $d_{xy}$-wave pairing $(\beta=\frac{\pi}{4})$. In addition, we still have $\delta=1$ for the $p$-wave SC and $\delta=0$ for the $d_{xy}$-wave SC.

Along the closed path in the middle WNSM in Fig. \ref{wws}(a), there are twice SAR processes. One is for the right-going electron which is specularly reflected as hole at $x=L$. The other is for the right-going hole which is specularly reflected as electron . Combing the acquired phases, the total phase obtained by the two SAR processes can be derived as given in Eqs. (\ref{sdqqc}) and (\ref{pdqqc}). Note, we have assumed $\theta>0$ for the $d_{xy}$-wave SC in the above discussions. The situation of $\theta<0$ can be considered in a similar way, which will not change the final results in Eqs. (\ref{sdqqc}) and (\ref{pdqqc}).

\subsection{SAR probability and conductance for the WNSM-WNSM-SC double junctions}
The wave function in the left WNSM ($x<0$) for the spin-up electron incidence is given by
\begin{eqnarray}
\psi_{LW}=\left(\begin{array}{c}
1\\
0\end{array}\right)e^{ik^{e\uparrow}_{x}x}
+a_{\downarrow}\left(\begin{array}{c}
0\\
1\end{array}\right)e^{-ik^{h\downarrow}_{x}x}
+b_{\uparrow}\left(\begin{array}{c}
1\\
0\end{array}\right)e^{-ik^{e\uparrow}_{x}x},
\end{eqnarray}
with $a_{\downarrow}$ and $b_{\uparrow}$ still the SAR and the normal reflection coefficients. The wave function in the middle WNSM ($0<x<L$) is
\begin{eqnarray}
\begin{split}
\psi_{MW}=f_{1}\left(\begin{array}{c}
1\\
0\end{array}\right)e^{ik^{e\uparrow}_{x}x}
+f_{2}\left(\begin{array}{c}
1\\
0\end{array}\right)e^{-ik^{e\uparrow}_{x}x}\\
+
f_{3}\left(\begin{array}{c}
0\\
1\end{array}\right)e^{ik^{h\downarrow}_{x}x}
+f_{4}\left(\begin{array}{c}
0\\
1\end{array}\right)e^{-ik^{h\downarrow}_{x}x}
\end{split}
\end{eqnarray}
where $f_{1}$ and $f_{2}$ are coefficients for the right-going electron and the left-going electron and $f_{3}$ and $f_{4}$ are coefficients for the right-going hole and the left-going hole. The wave function in SC ($x>L$) is written as
\begin{eqnarray}
\Psi_{S}=c_{\uparrow}\left(\begin{array}{c}
u_{+}e^{i\phi_{+}}\\
v_{+}\end{array}\right)e^{ik_{x}x}
+d_{\downarrow}\left(\begin{array}{c}
v_{-}e^{i\phi_{-}}\\
u_{-}\end{array}\right)e^{-ik_{x}x},
\end{eqnarray}
with $c_{\uparrow}$ and $d_{\downarrow}$ still the coefficients for quasiparticle transmissions. For the spin-down electron incidence, the wave functions can also be given in a similar way.

Using the following boundary conditions,
\begin{eqnarray}
\begin{split}
&\psi_{LW}|_{x=0}=\psi_{MW}|_{x=0},\\
&\psi_{MW}|_{x=L}=\Psi_{S}|_{x=L},\\
&\psi'_{MW}|_{x=0}-\psi'_{LW}|_{x=0}=\frac{2mV_1}{\hbar^2}\hat{\tau}_{z}\psi_{LW}|_{x=0},\\
&\Psi'_{S}|_{x=L}-\hat{\tau}_{z}\psi'_{MW}|_{x=L}=\frac{2mV_2}{\hbar^2}\psi_{MW}|_{x=L},
\end{split}
\end{eqnarray}
$a_{\downarrow}$ and $b_{\uparrow}$ will be solved. The similar boundary conditions for the spin-down electron incidence can lead to the solutions of $a_{\uparrow}$ and $b_{\downarrow}$. The probabilities for SAR and the normal reflection are defined according to Eq. (\ref{prob}). The conductance of the double junctions can also be written as Eqs. (\ref{fencond}) and (\ref{totalcond}).

\end{document}